\documentclass{ws-ijmpa}
\usepackage[super,compress]{cite}
\usepackage{graphicx}
\usepackage{slashed}
\usepackage{textcomp}

\makeatletter
\newbox\slashbox \setbox\slashbox=\hbox{$/$}
\newbox\Slashbox \setbox\Slashbox=\hbox{\large$/$}
\def\pFMslash#1{\setbox\@tempboxa=\hbox{$#1$}
	\@tempdima=0.5\wd\slashbox \advance\@tempdima 0.5\wd\@tempboxa
	\copy\slashbox \kern-\@tempdima \box\@tempboxa}
\def\pFMSlash#1{\setbox\@tempboxa=\hbox{$#1$}
	\@tempdima=0.5\wd\Slashbox \advance\@tempdima 0.5\wd\@tempboxa
	\copy\Slashbox \kern-\@tempdima \box\@tempboxa}

\def\miss#1{\ifmmode{/\mkern-11mu #1}\else{${/\mkern-11mu #1}$}\fi}
\makeatother

\begin{document}
\markboth{Y. FLORES-OREA AND J. J. TOSCANO}
{CPT violation induced neutrino charge radius$\cdots$}

%
\catchline{}{}{}{}{}
%

\title{Effects of CPT violation on the neutrino charge radius in the Standard Model Extension}

\author{Y. FLORES-OREA AND J. J. TOSCANO}

\address{Facultad de Ciencias F\'{\i}sico Matem\'aticas,
Benem\'erita Universidad Aut\'onoma de Puebla, Apartado Postal
1152, Puebla, Puebla, M\'exico. \\ yuridia.floreso@alumno.buap.mx, jtoscano@fcfm.buap.mx}

\maketitle

\begin{history}
\received{Day Month Year}
\revised{Day Month Year}
\end{history}

\begin{abstract}
CPT-Odd effects on the charge radius of the neutrino are studied in the context of the Standard Model Extension. Effects of CPT violation arising from the electroweak Yang-Mills sector, characterized by the Lorentz violation coefficients $(k_1)_\mu$ and $(k_2)_\mu$, which have positive mass units, are considered. In this context, we find that the $\bar{\nu}\nu \gamma$ vertex is induced at the tree level through the exchange of two $Z$ gauge bosons. Although strongly suppressed due to a factor of $\frac{1}{m^4_Z}$, the process has some characteristics that make it interesting. First, the vertex function $\Gamma_\mu$ is given by three independent gauge structures, which fulfill with a Ward identity of the way $q^\mu \Gamma_\mu=0$, which is characteristic of every neutral particle, and induce electromagnetic form factors that are independent on the gauge parameter $\xi$. At this level, besides the charge and anapole, two additional nonconventional form factors are generated. Second, the charge form factor $f_Q(q^2)$ has a piece that is energy dependent, which causes $f_Q(q^2=0)\neq 0$, so the electromagnetic properties cannot be defined for a real photon in the presence of CPT violation. Instead, the charge form factor and their associated charge radius are defined in the static limit, which is given by the kinematical conditions: $q^0=0$ and then $\mathbf{q}\to 0$. In the static limit, we find that $f_Q(q^0=0, \mathbf{q}\to 0)=0$ and a correction to the Standard Model prediction of the neutrino charge radius given by $<r^2_\nu >_{CPTV}=\frac{3c_{2W}}{2c^4_W}\left[\frac{k^2_2}{m^2_Z}+\frac{\mathbf{k_2^2}}{m^2_Z}\cos^2\theta_\gamma \right]\frac{1}{m^2_Z}$, with $\theta_\gamma$ the angle between the spatial vectors $\mathbf{q}$ and $\mathbf{k_2}$. Using a recently derived bound on the Lorentz violation coefficients $(k_i)_\mu$ along with some reasonable assumptions, we have obtained a small correction of $<r^2_\nu >_{CPTV} \leq 0.83 \times 10^{-51} cm^2$.

\end{abstract}



\section{Introduction}
\label{I}The electromagnetic properties of neutrinos have been the subject of great interest in the literature~\cite{Review}\,, meanly because they allow us to distinguish whether they are Dirac or Majorana neutrinos. In general, the electromagnetic vertex of a massive Dirac's neutrino is characterized by four real form factors, namely the charge, $F_Q(q^2)$, the magnetic dipole moment, $F_M(q^2)$, the electric dipole moment, $F_E(q^2)$, and the anapole moment, $F_A(q^2)$. Of course the charge form factor vanishes in an appropriate kinematic limit, but associated with it is the charge radius, which does not necessarily vanishes. It is worth mentioning that a massive Majorana neutrino cannot have neither magnetic nor electric dipole moment, although they could have flavour-off-diagonal transition magnetic and electric dipole moments. More generally, the electromagnetic properties of charged fermions are also characterized by these four form factors~\cite{AFF} . In quantum electrodynamics (QED), charged fermions can only have charge and magnetic moment form factors, which respect separately the discrete $C$ (charge conjugation), $P$ (parity), and $T$ (time reversal) transformations. At the one-loop level, the charge and magnetic dipole form factors are gauge independent, the former exhibiting  both ultraviolet (UV) and infrared (IR) divergences, while the latter is an unambiguous physical observable free of both types of divergences. The gauge independence of the charge and magnetic dipole moment in QED is a direct consequence of the Abelian character of the theory. However, this scenario is affected in a non-trivial way in the context of the electroweak theory, where non-Abelian interactions between chiral fermions lead to additional contributions to the charge and magnetic dipole moment predicted by QED, as well as the emergence of the electric dipole moment and the anapole moment. Under $C$, $P$, and $T$ transformations the electric dipole(anapole) form factors are even(odd), odd(odd), and odd(even), respectively. Hence, the electric dipole and anapole form factors are odd and even under $CP$ transformations, respectively. In the Standard Model (SM) the only source of $CP$ violations is the Cabbibo–Kobayashi–Maskawa (CKM) phase. In the SM, electric dipole moments of quarks are induced through this complex phase, but they are quite suppressed because they first arise up to the three-loop level~\cite{QEDM}. However, electric dipole moments of both leptons and quarks can be induced at lower orders in many well-motivated SM extensions.\\

The study of the electromagnetic properties of neutrinos in the SM has a long and interesting history. The first studies on these properties begin in the early 1970s simultaneously with the first tests of the SM at the one loop level~\cite{FSEPN}\,. In the following decade, the electromagnetic properties of neutrinos came under scrutiny in the context of general gauge theories~\cite{Shr} and more specifically within SM~\cite{LRZ,MRS,DSM}\,. As is well known, in the SM neutrinos have no mass, so they can only have charge and anapole form-factors. In this minimal version of the SM in which there are only left-handed neutrinos, the charge radius and the anapole coincide. In this context, the neutrino charge radius was calculated using linear~\cite{LRZ} and nonlinear~\cite{MRS} $R_\xi$ gauges. It was concluded that these quantities cannot be physical, since they depend on the gauge-fixing procedure and the gauge parameter $\xi$. In Ref.~\cite{DSM} an effective neutrino charge radius was proposed by extracting a gauge-independent  electromagnetic form factor from the low energy elastic scattering $\nu-l$ process, with $\nu$ and $l$ the Dirac's neutrino and charged lepton, respectively. The extraction of this charge radius is based on the gauge independence of the $S$-matrix element associated with the scattering process $\nu-l$, but has the disadvantage of being process dependent. Some time later, a gauge-independent and process-independent neutrino charge radius was obtained~\cite{NCRPT} through a rearrangement of diagrams using the Pinch Technique~\cite{PT}\,.

In this paper, we are interested in studying the implications of $CPT$ violation on the neutrino charge radius in the context of the Standard Model Extension (SME)~\cite{SME}\,, which is an effective theory that incorporates in a model-independent manner both Lorentz Violation (LV) and CPT violation (CPTV). In its minimal version (mSME), which is renormalizable in the Dyson's sense, the theory is given by a Lagrangian of the form:
\begin{equation}
{\cal L}={\cal L}_{SM}+\Delta {\cal L}\, ,
\end{equation}
where ${\cal L}_{SM}$ is the SM Lagrangian and $\Delta {\cal L}$ is formed by a sum of pieces of the form $T^{\mu_1\cdots \mu_n}{\cal O}_{\mu_1\cdots \mu_n}$. In this expression $T^{\mu_1\cdots \mu_n}$ is a constant Lorentz tensor, while ${\cal O}_{\mu_1\cdots \mu_n}$ is a Lorentz tensor containing the SM degrees of freedom, that is, it depends on the SM fields and is invariant under the SM gauge group. To understand how violations of Lorentz and CPT symmetries arise, we need the notions of observer Lorentz transformations (OLT) and particle Lorentz transformations (PLT). OLTs correspond to usual coordinate transformations. On the other hand, under PLT only the degrees of freedom (the SM fields) are transformed. Under OLT, both $T^{\mu_1 \cdots \mu_n}$ and ${\cal O}_{\mu_1 \cdots \mu_n}$ are transformed, so $\Delta {\cal L} $ is invariant under this type of transformations, as is the SM Lagrangian. However, under PLT ${\cal O}_{\mu_1 \cdots \mu_n}$ is transformed but $T^{\mu_1 \cdots \mu_n}$ is not, so $\Delta { \cal L}$ is not invariant under this type of transformations. The LV $\Delta { \cal L}$ Lagrangian naturally appears  formed of two pieces, one CPT-Even  and other CPT-Odd. It turns out that the LV coefficients $T^{\mu_1\cdots \mu_n}$ associated with the CPT-Even terms are dimensionless, while those associated with the CPT-Odd terms have positive mass dimension~\footnote{Dimensionless CPT-Odd Lorentz coefficients can arise in the context of the QED extension, but cannot be obtained as a reduction of the SME because they are incompatible with electroweak symmetry~\cite{SME}.}. Since the theory is renormalizable, CPT-Even interactions have canonical dimension four, while CPT-Odd interactions have dimension three. In general, an effective field theory (nonrenormalizable in the Dyson's sense) have three types of parameters, namely relevant parameters, which have positive mass units, marginal parameters, which are dimensionless, and irrelevant parameters, which have negative mass units. Since the mSME is renormalizable in the Dyson's sense, it only has relevant and marginal parameters. To appreciate the relative importance of the relevant ($M_i$) and irrelevant ($\Lambda^{-1}_i$) parameters, consider a given process of energy $E$ ($E$ can be the energy of a scattering process or the mass of a decaying particle). Due to dimensional considerations, the contribution of the relevant parameters will be of the form $\frac{M_i}{E}$, while the contribution of the irrelevant parameters will be proportional to $\frac{E}{\Lambda_i}$. This means that at low energies ($E<M_i$ and $E\ll \Lambda_i$), the contribution of the relevant parameters dominates but that of the irrelevant parameters is subdominant. The opposite occurs at high energies ($M_i\ll E< \Lambda_i$), since in this case the contribution of the irrelevant parameters dominates. The marginal parameters would be important at all energy scales.

As already commented, in this work we will investigate a possible neutrino charge radius induced by CPTV. We focus on the CPTV electroweak Yang-Mills sector of the mSME, since it can induce the neutrino charge radius at the tree level. This CPTV sector is characterized by two LV coefficients $(k_i)_\mu$ ($i=1,2$), which have positive mass dimension and transform as vectors under OLT. As we will see later, the charge radius of the neutrino is generated up to second order in the LV coefficients, that is, it is proportional to the Lorentz-scalar $k^2_i=(k_i)_\mu (k_i)^\mu $. Very recently, in a paper for one of us~\cite{OP1} , it was shown that these LV coefficients are quite sensitive to low-energy processes, which allowed us to obtain a bound for $k^2_i$ from the experimental measurement of the magnetic dipole moment of the electron. We will use this limit to estimate the size of the correction to the SM prediction of the neutrino charge radius.

The rest of the paper has been organized as follows. In Sec.~\ref{YM} the gauge structure of the CPTV electroweak Yang-Mills sector is discussed. The Feynman rules needed for the calculation will be derived. Sec.~\ref{Cal} is devoted to present the calculation of the neutrino charge radius. In Sec.~\ref{D} we discuss our results. Finally, in Sec.~\ref{C} the conclusions are presented.

\section{Electroweak Yang-Mills sector with CPT violation}
\label{YM}The electroweak Yang-Mills sector of the SME that is odd under CPT transformations is given by~\cite{SME,OP1,CNP}:
\begin{eqnarray}
	\label{LW}
	{\cal L}^{CPT-Odd}_{W}&=&\frac{1}{2}(k_2)_\lambda \epsilon^{\lambda \rho \mu \nu}Tr\left(W_\rho W_{\mu \nu}+\frac{2}{3}igW_\rho W_\mu W_\nu \right)\, , \\
	\label{LB}
	{\cal L}^{CPT-Odd}_{B}&=&\frac{1}{2}(k_1)_\lambda  \epsilon^{\lambda \rho \mu \nu}B_\rho B_{\mu \nu}\, ,
\end{eqnarray}
where $W_\mu=\frac{\sigma^i}{2}W^i_\mu$ and $W_{\mu \nu}=\frac{\sigma^i}{2}W^i_{\mu,\nu}$ are the gauge field and curvatures of the $SU_L(2)$ gauge group, whereas $B_\mu$ and $B_{\mu \nu}$ are the gauge field and curvature of the $U_Y(1)$ gauge group. In addition, the $\frac{\sigma^i}{2}$ metrices are the generators of $SU_L(2)$ in the fundamental representation and $g$ is its coupling constant. It should be noted that the above Lagrangians are gauge invariant up to a boundary term.

Moving from the  electroweak group $SU_L(2)\times U_Y(1)$  description to the electromagnetic group $U_Q(1)$ perspective through the usual mapping of the $(W^1, W^2, W ^3, B)$ fields to the $(W^{-},W^{+}, Z, A)$ fields, we have
\begin{eqnarray}
	\label{LWW}
 {\cal L}^{CPT-Odd}_{W}=\frac{1}{4}(k_2)_\lambda \epsilon^{\lambda \rho \mu \nu}\left(W^+_\rho \hat{W}^-_{\mu \nu}+W^-_\rho \hat{W}^+_{\mu \nu}+
	W^3_\rho W^3_{\mu \nu}-\frac{2ig}{3}W^3_{[\rho} W^+_\nu W^+_{\mu]} \right)\, ,
\end{eqnarray}
\begin{equation}
	\label{LBB}
	{\cal L}^{CPT-Odd}_{B}=\frac{1}{2}(k_1)_\lambda \epsilon^{\lambda \rho \mu \nu}\left[c^2_W A_\rho F_{\mu \nu}+s^2_W Z_\rho Z_{\mu \nu}-s_Wc_W\left(A_\rho Z_{\mu \nu}+Z_\rho F_{\mu \nu} \right) \right]\, .
\end{equation}
In the above expressions,
\begin{eqnarray}
	\hat{W}^+_{\mu \nu}&=&W^+_{\mu \nu} +ig(W^+_\mu W^3_\nu-W^+_\nu W^3_\mu)\, , \\
	\hat{W}^-_{\mu \nu}&=&\left(\hat{W}^+_{\mu \nu}\right)^\dag\, , \\
	W^3_{\mu \nu}&=&s_WF_{\mu \nu}+c_W Z_{\mu \nu}+ig\left(W^-_\mu W^+_\nu -W^+_\mu W^-_\nu \right)\, ,
\end{eqnarray}
where $W^+_{\mu \nu}=\partial_\mu W^+_\nu-\partial_\nu W^+_\mu$, $W^3_\mu=c_WZ_\mu +s_WA_\mu$, $Z_{\mu \nu}=\partial_\mu Z_\nu-\partial_\nu Z_\mu$, and $F_{\mu \nu}=\partial_\mu A_\nu-\partial_\nu A_\mu$. In Eq.(\ref{LWW}), the symbol $[\rho \nu \mu]$ represents the sum of cyclic permutations in the indices $\rho$, $\nu$ and $\mu$. In addition, $s_W(c_W)$ stands for sine(cosine) of the weak angle $\theta_W$. The above Lagrangians induce quadratic and cubic couplings between the various electroweak gauge bosons, all of which are treated perturbatively. We focus on the quadratic interactions between the photon and the $Z$ gauge boson, which are given by the following Lagrangians~\cite{OP1}
\begin{eqnarray}
	\label{CFJ}
	{\cal L}^{CPT-Odd}_{AF}&=&\frac{1}{4}(k_{AF})_\lambda \epsilon^{\lambda\rho \mu \nu}A_\rho F_{\mu \nu} \, , \\
	{\cal L}^{CPT-Odd}_{ZZ}&=&\frac{1}{4}(k_{ZZ})_\lambda \epsilon^{\lambda\rho \mu \nu}Z_\rho Z_{\mu \nu} \, , \\
	{\cal L}^{CPT-Odd}_{AZ}&=&\frac{1}{4}(k_{AZ})_\lambda \epsilon^{\lambda\rho \mu \nu}\left(A_\rho Z_{\mu \nu}+Z_\rho F_{\mu \nu}\right) \, ,
\end{eqnarray}
where
\begin{eqnarray}
	\label{CLAA}
	k_{AF}&=&2c^2_Wk_1+s^2_Wk_2\, , \\
	\label{CLZZ}
	k_{ZZ}&=&2s^2_Wk_1+c^2_Wk_2\, , \\
	\label{CLAZ}
	k_{AZ}&=&s_Wc_W(k_2-2k_1)\, .
\end{eqnarray}
The LV coefficient $k_{AF}$ given by Eq.(\ref{CLAA}), which is associated with the Carroll-Field-Jackiw's Lagrangian~\cite{CFJ}, has been an important focus of attention in the literature. In this pioneering work, data on cosmological birefringence were used to impose upper bounds of order of $10^{- 25} \, GeV$ and $10^{-42}\, GeV$, for a Lorentz coefficient of the form $(k_{AF})^\alpha=(k,\vec{0})$ and for a timelike $(k_{AF})^\alpha$, respectively. Recently, stronger upper bounds have been obtained from cosmic microwave background searches~\cite{Caloni}. In this work, an upper bound on the parameter $k^{(3)}_{(V)00}=-\sqrt{4\pi}(k_{AF})^0$ of $|k^{(3)}_{(V)00}|<1.54\times 10^{-44}\, GeV$ has been obtained. In the same work, a bound of $|\mathbf{k}_{AF}|<7.4\times 10^{-45}\, GeV$ on the space components of $(k_{AF})_\mu$ was obtained. Bounds on these parameters derived in previous literature are collected in table D15 of  reference~\cite{TK}.

\begin{figure}
\centering\includegraphics[width=3.0in]{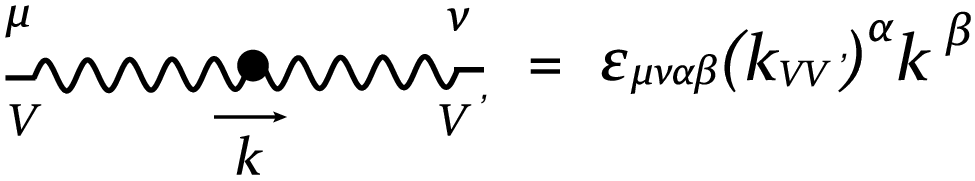}
\caption{\label{fig1}Feynman rule for the quadratic couplings $ZZ$ and $AZ$.}
\end{figure}

\begin{figure}
\centering\includegraphics[width=3.0in]{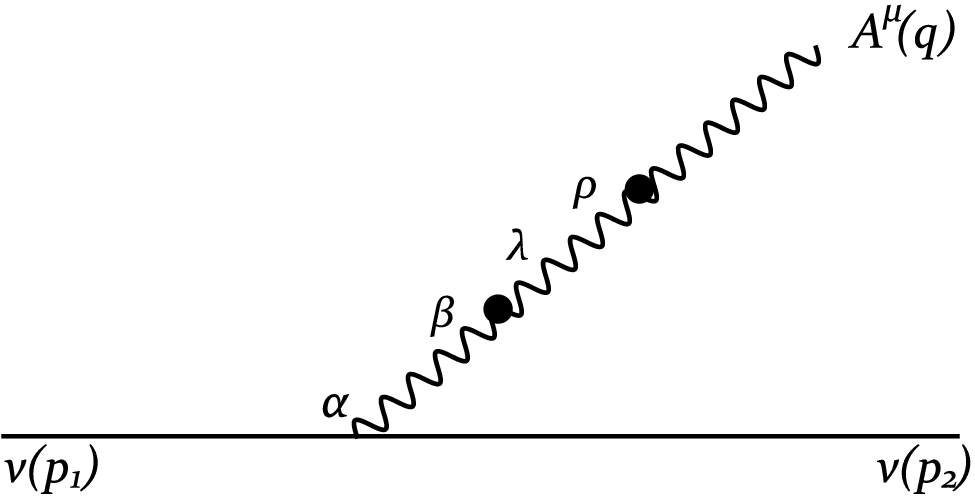}
\caption{\label{Fig2}Feynman diagram contributing to the $\bar{\nu}\nu \gamma$ vertex at the tree level.}
\end{figure}

Because of the very strong bounds on the Lorentz coefficient $(k _{AF})_\mu$, it makes sense to assume that this parameter is, for all practical purposes, equal to zero. This in turns implies from the relation given by Eq.(\ref{CLAA}) that the Lorentz coefficients $(k_1)_\mu$ and $(k_2)_\mu$ are collinear, that is,
\begin{equation}
	\label{cr}
	(k_1)_\mu=-\frac{1}{2}\frac{s^2_W}{c^2_W}(k_2)_\mu \, .
\end{equation}
Working with this assumption, the electromagnetic neutrino $\bar{\nu}\nu \gamma$ vertex is generated at the tree level (see Fig.~\ref{Fig2}) by the SM $\bar{\nu}\nu Z$ vertex and the CPTV vertices $ZZ$ and $ZA$. The Feynman rules for the quadratic couplings are given in Fig.~\ref{fig1}.

\section{The calculation}
\label{Cal}The contribution to the charge radius of the neutrino is given by diagram in Fig.~\ref{Fig2}~\footnote{A contribution of $O(k_i)$ to the coupling $\bar{\nu}\nu \gamma$ can be induced at the tree level by the vertex $ZA$, but it does not contribute to the charge radius.}. The contribution is mediated by two virtual $Z$ gauge bosons. In the $R_\xi$ gauge, the propagator of this gauge boson is given by:
\begin{equation}
\Delta^{\alpha \beta}=\frac{-i}{q^2-m^2_Z}\left(g^{\alpha \beta}-(1-\xi)\frac{q^\alpha q^\beta}{q^2-\xi m^2_Z}\right)\, ,
\end{equation}
where $\xi$ is the gauge parameter. On the other hand, the vertices $ZZ$ and $ZA$ are given by $\Gamma_{\beta \lambda}=\epsilon_{\beta \lambda \omega \sigma}(k_{ZZ})^\omega q^\sigma$ and $\Gamma_{\rho \mu}=\epsilon_{\rho \mu \zeta \tau}(k_{AZ})^\zeta q^\tau$, so the contractions among the two $Z$ propagators and these two vertices lead to a result of the form:
\begin{equation}
\Delta^{\alpha \beta} \Gamma_{\beta \lambda} \Delta^{\lambda \rho}\Gamma_{\rho \mu}=-\frac{g^{\alpha \beta}g^{\lambda \rho}}{(q^2-m^2_Z)^2}\Gamma_{\beta \lambda} \Gamma_{\rho \mu}\, ,
\end{equation}
which is independent of the gauge parameter $\xi$. This means that the electromagnetic vertex $\bar{\nu}\nu \gamma$ arising from CPTV is gauge-independent. Using the collinear relation (\ref{cr}), the amplitude with the photon off-shell can be written as follows:
\begin{equation}
\label{AI}
{\cal M}_\mu=ie\, \bar{u}(p_2)\Gamma_\mu (q,k_1) u(p_1)\, ,
\end{equation}
 with $\Gamma_\mu(q,k_1)$ the vertex function given by
 \begin{equation}
 \label{VF}
 \Gamma_\mu(q,k_1)=\frac{2 c_{2W}}{s^4_W}\frac{1}{(q^2-m^2_Z)^2}\left(\gamma^\alpha P_L\right)\left[k^2_1P_{1 \alpha \mu}+\left(k_1\cdot q\right)P_{2 \alpha \mu}+P_{3 \alpha \mu}\right]\, ,
 \end{equation}
 where $c_{2W}=c^2_W-s^2_W$ and $P_L=\frac{1}{2}(1-\gamma_5)$ is the left-handed projector. In addition,
 \begin{eqnarray}
 P_{1\alpha \mu}&=&q^2g_{\alpha \mu}-q_\alpha q_\mu \, , \\
 P_{2\alpha \mu}&=&k_{1\mu}q_\alpha -(k_1\cdot q)g_{\alpha \mu}\, ,\\
 P_{3\alpha \mu}&=&(k_1\cdot q)k_{1\alpha}q_\mu-q^2 k_{1 \alpha}k_{1\mu}\, .
 \end{eqnarray}
 The above $P_{i\alpha \mu}$ ($i=1,2,3$) tensors represent electromagnetic gauge structures in the sense that they satisfy the Ward identities:
 \begin{equation}
 q^\mu P_{i\alpha \mu}=0\, .
 \end{equation}
 The Ward identity $q^\mu \Gamma_\mu=0$ is consistent with the fact that the neutrino is a neutral particle, since if it were a charged particle, the difference between the vertex functions that characterize the external legs should appear on the right-hand side. In other words, this coupling is not generated through the covariant electromagnetic derivative, as it must be for a neutral particle.

 It is easy to see that the $P_{1\alpha \mu}$ gauge structure induce both the charge and anapole form factors of the neutrino. In fact, from
\begin{equation}
\label{Cha}
(\gamma^\alpha P_L)k^2_1 P_{1\alpha \mu}=k^2_1\left(q^2\gamma_\mu-\pFMSlash{q} q_\mu\right)P_L\, ,
\end{equation}
we can recognize from the vector and axial-vector parts of this expression the charge and anapole Lorentz structures, respectively. This gauge structure arises from an interaction of canonical dimension six of the form
\begin{equation}
\sim k^2_1 \, \bar{\nu}\gamma_\alpha P_L \nu \,  \partial_\beta F^{\beta \alpha}\, ,
\end{equation}
where $\nu(x)$ is the spinor field associated with the neutrino.

On the other hand, the $P_{2\alpha \mu}$ gauge structure leads to
\begin{equation}
\label{P2}
(\gamma^\alpha P_L )(k_1\cdot q) P_{2\alpha \mu}=-(k_1\cdot q)\left((k_1\cdot q)\gamma_\mu-\pFMSlash{q}k_{1\mu}\right)P_L\, .
\end{equation}
Note that the Lorentz structure of this expression has a great resemblance to the one given in (\ref{Cha}). However, it does not vanish on-shell. As we will see below, the vector part of this expression induces an energy-dependent contribution to the charge form factor. This gauge structure arises from a dimension-six interaction of the form:
\begin{equation}
\sim k_{1\lambda}k_{1\beta}\left(\bar{\nu}\gamma_\alpha P_L\nu\right)\partial^\lambda F^{\beta \alpha}\, .
\end{equation}

As far as the $P_{3\alpha \mu}$ gauge structure is concerned, it leads to
\begin{equation}
\label{P3}
(\gamma^\alpha P_L )P_{3\alpha \mu}=\left[(k_1\cdot q)q_\mu-q^2 k_{1\mu}\right]\pFMSlash{k_1}P_L \, .
\end{equation}
Note that this amplitude vanishes on-shell. This gauge structure does not contribute to any of the conventional electromagnetic properties of the neutrino. This term arises from a Lagrangian of the form:
\begin{equation}
\sim k_{1\alpha}k_{1\lambda}\bar{\nu}\gamma^\alpha P_L \nu \partial_\rho F^{\lambda \rho}\, .
\end{equation}

From the above results, we can see that the vertex function can be expressed in terms of four electromagnetic gauge structures as follows:
\begin{eqnarray}
\label{VF1}
\Gamma_\mu(q,k_1)&=&f_Q(q,k_1)\, \gamma_\mu +f_A(q,k_1) \left(q^2\gamma_\mu- \pFMSlash{q}q_\mu \right)\gamma_5\, \nonumber \\
&+&f\, \left[(k_1\cdot q)\left((k_1\cdot q)\gamma_\mu-\pFMSlash{q}k_{1\mu}\right)\gamma_5\right]\, \nonumber \\
&+&2f\left[(k_1\cdot q)q_\mu-q^2 k_{1\mu}\right]\pFMSlash{k_1}P_L\, ,
\end{eqnarray}
where the charge, anapole, and nonconventional form factors are respectively given by
\begin{eqnarray}
\label{CFF}
f_Q(q,k_1)&=&\frac{c_{2W}}{s^4_W}\, \frac{k^\alpha_1 k^\beta_1}{(q^2-m^2_Z)^2}\left(q^2g_{\alpha \beta}-q_\alpha q_\beta \right)\, \\
\label{AFF}
f_A(q^2,k_1^2)&=&-\frac{c_{2W}}{s^4_W}\, \frac{k^2_1}{(q^2-m^2_Z)^2}\, ,\\
f&=&\frac{c_{2W}}{s^4_W}\frac{1}{(q^2-m^2_Z)^2}\, .
\end{eqnarray}
Note that the vertex function (\ref{VF1}) contains, in addition to the charge and the anapole, two nonconventional gauge structures that depend on the LV coefficient $k_{1\mu}$. Also, note that $f_A$ is invariant under both OLT and PLT, but $f_Q$ is only invariant under OLT since it depends on the energy through the term $(k_1\cdot q)^2$ (in the on-shell case $q^2=0$). The structure of the charge form factor is striking, since it can be written as the contraction $k_1^\alpha k_1^\beta \pi_{\alpha \beta}$, with $\pi_{\alpha \beta}$ the gauge structure that characterizes the photon self-energy. In the next section, we study the structure of the charge form factor and its associated charge radius.

\section{Discussion}
\label{D} We now move on to study the charge radius associated with the charge form factor $f_Q(q,k_1)$. In conventional theories~\footnote{Quantum Field Theories in the vacuum, which respect the Lorentz and $CPT$ symmetries.}, the charge radius is defined by the second term of the Taylor expansion around $q^2=0$,
\begin{equation}
\label{TS1}
f_Q(q^2)=f_Q(0)+q^2\,\left\{ \frac{df_Q}{dq^2}\right\}_{q^2=0}+\cdots\, .
\end{equation}
In the case of the neutrino, $f_Q(0)=0$ and the charge radius is defined by
\begin{equation}
\label{rc1}
<r^2_\nu>=6\left\{\frac{df_Q}{dq^2}\right\}_{q^2=0}\, .
\end{equation}
However, from the structure of $f_Q(q,k_1)$ given in (\ref{CFF}), we immediately see that the charge defined for a photon on the mass shell is non-zero. This due to the presence of the energy-dependent factor $(k_1\cdot q)^2$, which, for the photon on-shell, is equal to $[q_0(k^0_1-|\mathbf{k_1}|cos\theta_\gamma)]^2$, with $\theta_\gamma$ the angle between the $\mathbf{k_1}$ and $\mathbf{q}$ spatial vectors. This unexpected result means that we need to be more careful about how we define the charge. A more precise form of defining the electric charge~\cite{NP} consists in assuming the same energy for the two neutrinos, that is, $p_1=(E,\mathbf{p_1})$ and $p_2=(E,\mathbf{p_2})$, so that $q^0=0$ and $\mathbf{q}=\mathbf{p_1}-\mathbf{p_2}$. The limiting procedure of taking $q^0=0$ and then $\mathbf{q} \to 0$ is known as the static limit. We will adopt this procedure to define the charge and the charge radius.

In the static limit,
\begin{eqnarray}
f_Q(q^0=0,\mathbf{q},k_1)&=&-\frac{c_{2W}}{s^4_W}\frac{1}{(\mathbf{q}^2+m_Z^2)^2}\left[k^2_1\mathbf{q}^2+(\mathbf{q}\cdot \mathbf{k_1)^2}\right]\, \nonumber\\
&=&-\frac{c_{2W}}{s^4_W}\left[(k^0_1)^2-(\mathbf{k_1})^2\sin^2\theta_\gamma\right]\frac{\mathbf{q}^2}{(\mathbf{q}^2+m_Z^2)^2}\, ,
\end{eqnarray}
and clearly $f_Q(q^0=0,\mathbf{q}\to 0,k_1)=0$. Then, the associated charge radius is calculated as follows:
\begin{equation}
<r^2_\nu >_{CPTV}=\left\{6\frac{df_Q(q^0=0,\mathbf{q},k_1)}{d\mathbf{q}^2}\right\}_{\mathbf{q}^2=0}\, ,
\end{equation}
which leads to
\begin{eqnarray}
\label{chr1}
<r^2_\nu >_{CPTV}&=&-\frac{6c_{2W}}{s^4_W}\frac{(k^0_1)^2-(\mathbf{k_1})^2\sin^2\theta_\gamma}{m^4_Z}\, \nonumber \\
&=&-\frac{3c_{2W}}{2c^4_W}\frac{(k^0_2)^2-(\mathbf{k_2})^2\sin^2\theta_\gamma}{m^4_Z}\, ,
\end{eqnarray}
where in the last step the collinear condition (\ref{cr}) was used. Note that the charge radius of the neutrino can be, in general, negative. From Eq.(\ref{AFF}) we see that in the static limit, the charge radius and anapole form factor satisfy the following relation:
\begin{equation}
<r^2_\nu >_{CPTV}=6 f_A(q^0=0,\mathbf{q}\to 0)+\frac{3c_{2W}}{2 c^4_W}\left(\frac{\mathbf{k_1}^2}{m^4_Z}\cos^2\theta_\gamma\right)\, ,
\end{equation}
which reduces to the well known SM relation if the energy dependent term in the charge form factor is ignored.

In terms of the OLT invariant $k^2_2=(k_2)_\mu (k_2)^\mu=(k^0_2)^2-\mathbf{k_2}^2$, we can rewrite the charge radius given in (\ref{chr1}) as follows
\begin{equation}
<r^2_\nu >_{CPTV}=-\frac{3c_{2W}}{2c^4_W}\left[\frac{k^2_2}{m^2_Z}+\frac{\mathbf{k_2^2}}{m^2_Z}\cos^2\theta_\gamma \right]\frac{1}{m^2_Z}\, .
\end{equation}
 Then we can use the bound $|k^2_2|<4.36\times 10^{-10} m^2_e$ obtained~\cite{OP1} from the experimental value for the electron magnetic dipole moment to estimate $<r^2_\nu >_{SME}$. Since this bound is valid for both timelike or spacelike $k_2$, the CPT-Odd contribution to the charge radius can be positive or negative. Assuming that $\mathbf{k_2}^2\cos^2\theta_\gamma \leq k^2_2$, we find that the CPT violation effects arising from the electroweak gauge sector on the neutrino charge radius must be, in absolute value, smaller than $0.83\times 10^{-51} cm^2$. Such contribution must be less than this number because the bound on $k^2_2$ is an upper bound~\cite{OP1}.

 In the SME, the above result is the CPTV correction to the SM prediction of the neutrino charge radius, so that including these effects, we have
 \begin{equation}
 <r^2_\nu >_{SME}=<r^2_\nu >_{SM}+<r^2_\nu >_{CPTV}\, ,
 \end{equation}
 where the SM prediction is given by ~\cite{NCRPT}
 \begin{equation}
 <r^2_\nu >_{SM}=\frac{G_F}{4 \sqrt{2}\pi^2}\left[3-2\log\left(\frac{m^2_l}{m^2_W} \right) \right]\, ,
 \end{equation}
 where $G_F$ is the Fermi constant.

 This extremely small contribution of CPTV to the neutrino charge radius, which is eighteen orders of magnitude smaller than the value predicted by SM, is due in part to the fact that the process is completely dominated by the exchange of two virtual $Z$ gauge bosons. However, this scenario can change radically if the one-loop effects of $O(k^2_i)$  in the LV coefficients are considered, since, as demonstrated recently for the case of the magnetic moment of charged leptons\cite{OP1}, other natural energy scales come into play at this order, namely the masses $m_l$ of the charged leptons. At this order, the presence of these new energy scales together with the fact that the LV coefficients belong to the category of relevant parameters, conspire to produce a non-decoupling effect (in the sense of $m_Z \gg m_l$) that gives rise to a contribution greater than that at the tree level by at least a factor of $\left(\frac{m^2_Z}{m^2_l}\right)$. We will postpone this study for a future research.

\section{Conclusions}
\label{C}In this work, effects of CPTV on the charge radius of the neutrino arising from the electroweak gauge sector of the SME were studied. It was found that the neutrino charge radius is induced through $O(k^2_i)$ effects of the LV coefficientes at the tree level via the interchange of two virtual $Z$ gauge bosons. It was shown that, in addition to the gauge structure that gives rise to the charge and anapole  moment form factors, two new unconventional gauge structures are generated. We find that all of these electromagnetic properties of the neutrino are physical in the sense that they are independent of the gauge parameter $\xi$. We find that while the charge form factor depends on energy, the corresponding form factor associated with the anapole does not. The anapole form factor is found to be well defined for a real photon with a value independent on the energy given by $f _A(q^2=0,k^2_1)=-\frac{c_{2W}}{s^4 _W}\frac {k^2_1}{m^4_Z}$. However, we find that the charge form factor does not vanishes on the mass shell, which is a consequence of the fact that it depends on the energy through a term of the form $(q\cdot k_1)^2$. Using the static limit, which is defined by the conditions $q^0=0$, $\mathbf{q}\to 0$, we find that the charge form factor vanishes, while the anapole form factor coincides with the one obtained in the on-shell case. In the static limit, we obtain a CPTV correction to the charge radius of the neutrino given by $<r^2_\nu >_{CPTV}=-\frac{3c_{2W}}{2c^4_W}\left[\frac{k^2_2}{m^2_Z}+\frac{\mathbf{k_2^2}}{m^2_Z}\cos^2\theta_\gamma \right]\frac{1}{m^2_Z}$. Using a recently bound on $|k^2_2|$ obtained from experimental results on the electron magnetic dipole moment, we find that $|<r^2_\nu >_{CPTV}|\leq 0.83\times 10^{-51} cm^2$, which is eighteen orders of magnitude smaller than the SM prediction. This extremely small correction is due to the strong suppression proportional to $\frac{1}{m^4_Z}$ arising from the exchange of two virtual $Z$ gauge bosons. However, this correction is expected to be substantially improved due to effects of $O(k^2_i)$ at the one loop level, since at this order the masses of the charged leptons come into play along with non-decoupling effects, as has been shown to occur in the case of the electromagnetic properties of charged leptons~\cite{OP1}. The study of this much more complicated but interesting problem is being carried out.

\section*{Acknowledgments}
We acknowledge financial support from Consejo Nacional de Humanidades, Ciencia y Tecnolog\'\i a (CONACHYT). J.J.T. also acknowledge financial support from CONAHCYT through Sistema Nacional de Investigadoras e investigadores (SNII).

\end{document}